# Enhancing Prostate Cancer Segmentation for Multi-Domain Generalization using a novel Parallel-Route Coherent Mixup Regularization Training

Josiah Simeth, Sudharsan Madhavan, Victoria Brennan, Sean McBride, James Janopaul-Naylor, Justin Haseltine, Daniel Gorovets, Himanshu Nagar, Neelam Tyagi and Harini Veeraraghavan

Memorial Sloan Kettering Cancer Center, 1275 York Avenue, New York, NY 10065

## ABSTRACT

**Objective:** Magnetic resonance imaging (MRI) guided adaptive radiotherapy (MRgART) for prostate cancer (PCa) can precisely target tumors while sparing organs of unnecessary radiation. Daily treatment adaptation requires accurate segmentation of tumors and organs. Manual delineation by experts can be time and cost prohibitive. Deep learning segmentation methods have limited success when applied to datasets distinct from training, hampering generalizability and adoption of MRgART.

**Approach:** We developed a novel parallel route coherent feature mixup (PaRC-mix) training approach for single source to multi-domain generalization. PaRC-mix creates feature augmentations at multiple network layers through linear combination of features extracted from different training samples in a batch. PaRC-mix training was implemented on two different deep and residually connected networks, a multiple resolution residual network (MRRN) and UNet++ to segment PCa dominant intraprostatic lesions from apparent diffusion coefficient images. Models were trained on 2,029 samples from 3.0T GE MRI and tested on 1,547 PCa samples from 5 datasets acquired using 3T Siemens, 3T Philips, and 1.5T Elekta Unity MR-Linac scanners.

**Main Results**: PaRC-mix training led to significantly more accurate tumor detection ($p < 0.001$) and segmentation ($p < 0.001$) for both networks compared to training without mixup as well as input-mix training. PaRC-mix also achieved better recall to precision tradeoff than mixup applied only on the network backbone or input-mixup. Using on a normalized composite DSC, HD95, and MSD score the accuracy gap between aggressive and non-aggressive lesions decreased from 21.1 and 19.5 for MRRN and UNet++ models trained without mixup to 5.2 and 7.9 with same models trained with PaRC-mix. Finally, PaRC-mix derived tumor dose conformity index was strongly correlated ($\rho = 0.875$) to metrics derived from clinical segmentations.

**Significance**: This paper presents an easy to implement network agnostic approach to feature augmentation in multi-stream networks that enhances generalizability for the difficult problem of prostate cancer lesion segmentation.

## INTRODUCTION

Prostate cancer (PCa) is a common male cancer worldwide. Radiation treatment (RT) is a highly effective treatment for PCa. Magnetic resonance imaging (MRI) guided radiation treatment is a new treatment modality where an onboard MRI imager is available with the linear accelerator to more precisely localize tumors, allowing daily treatment adaptations and enhance therapeutic outcomes (Dearnaley *et al.*, 2014; Sritharan and Tree, 2022). Randomized clinical trials have shown that boosting the radiation dose to intra-prostatic PCa lesions can reduce risk of local recurrence and improve biochemical recurrence free survival for patients with PCa (Steenbergen *et al.*, 2015; Murray *et al.*, 2020; Yasar *et al.*, 2024). However, this treatment requires highly accurate segmentation of PCa lesions on all RT fractions for precise radiation delivery. Manual delineation of PCa lesions is very time consuming, subject to inter-rater variabilities, and requires dedicated experts for daily adaptation, which is both resource intensive and impractical in most hospital settings (M. Y. Chen *et al.*, 2020; Jeganathan *et al.*, 2023; Salgues *et al.*, 2024).

Deep learning (DL) methods, both commercial and academic typically focus on segmenting the whole prostate gland and have demonstrated high accuracies (Elguindi *et al.*, 2019; Soerensen *et al.*, 2021). Unlike prostate gland, segmentation of PCa dominant intraprostatic lesions is challenging and prone to inter-rater delineation variability for various reasons including poor tumor-tissue differentiation, heterogeneous appearances and use of multi-modal imaging, small lesion volumes (often under 2 $cm^3$), and inconsistent appearance across the peripheral, central, and transition zones due to the complex anatomy of the prostate gland (M. Y. Chen *et al.*, 2020; Dai *et al.*, 2020; Liechti *et al.*, 2020). Hence, consistently accurate localization and delineation of PCa is essential for accurate delivery of biologically effective dose for safer and effective adaptive radiation treatments.

Functional image sequences such as apparent diffusion coefficient (ADC) MRI capture tumor specific information, and have shown to enhance segmentation accuracies (Esen *et al.*, 2013). Although ADC is a semi-quantitative imaging modality, it is impacted by differences in MRI acquisitions and scanning protocols, depicting considerable variations with respect to training data. DL models perform well on datasets that follow same distribution or "in-distribution" (ID) as training but deteriorate in performance when applied to out of distribution (OOD) datasets. OOD datasets commonly occur in most clinical scenarios even within same institution, which makes applying models to different institution datasets a challenging, unsolved problem.

Multi-source domain adaptation mitigates accuracy degradation on OOD datasets by including diverse multi-institution datasets into training. This is often not practical due patient privacy concerns and difficulty in obtaining manually delineated datasets for each institution (Dai *et al.*, 2020; Rodrigues *et al.*, 2024). Another set of approaches employ generative adversarial networks (GAN) to synthesize different domain datasets and increase data diversity (Jiang *et al.*, 2018; Jiang and Veeraraghavan, 2020; Zhu, Du and Yan, 2020; Chiou *et al.*, 2021). Whereas the latter approach alleviates the need for labeled multi-domain datasets, it still requires unlabeled multi-institutional datasets for GAN training. Models trained on data from a single source institution and then refined through few-shot fine tuning have shown to be effective for prostate gland segmentation from ultrasound images. (Vesal *et al.*, 2022) Alternatively, domain invariant features such as organ shapes have shown to be effective to achieve multi-domain generalization (Liu *et al.*, 2022). Whereas geometry is a useful feature for organs, cancerous lesions that depict considerable appearance and shape changes even in same patient undergoing treatment are less effective as a feature to inform segmentation.

A clinically practical approach to domain generalization is to train on a single source while enabling applicability to multiple target domain datasets by training models to extract domain-invariant image features (Liu *et al.*, 2022). As the primary source of data variation occurs from MR signal intensity distribution differences, prior works used simple data augmentation such as color or intensity jittering (Volpi *et al.*, 2018), as well as geometry constrained priors to enhance generalizability, albeit with limited success (Liu *et al.*, 2022). Our approach overcomes these challenges by using a novel approach that creates augmented features to model variations in the data and achieve multi-domain generalization with single-source training. Our method called **Pa**rallel **R**oute **C**oherent **mix**up or (PaRC-mix) is based on and extends manifold mixup methods, that extract variations in features extract at either single randomly chosen network layer (Verma *et al.*, 2019) as well as different layers (encoder versus bottleneck versus decoder) of network for segmentation (Jung *et al.*, 2019). PaRC-mix differs from the aforementioned methods in that it computes multiple feature augmentations in all the network layers thus increasing the range of variations in the features processed by the network.

Our key contribution is a multi-feature layer augmentation method called PaRC-mix that employs a coherent feature mixup strategy to create a rich set of feature variations by leveraging both network backbone and intermediate layers such as residual feature streams and deep skip layers. This paper builds upon our recent work (Simeth *et al.*, 2026). We extend that work by (i) expanding the analysis to additional OOD datasets, evaluating generalization on 4 different multi-institutional datasets with 3 different scanner manufacturers comprising 1,428 PCa samples, (ii) assessing performance via surface alignment (HD95 and MSD) in addition to DSC and detection statistics, (iii) using t-SNE feature clustering methods to better understand why generalization occurs with PaRC-mix, and (iv) validating the method as a network agnostic approach by implementing and testing the performance using both UNet++ (Zhou *et al.*, 2018) and the deep and residually connected network called multiple resolution residual network (MRRN). (Jiang *et al.*, 2019)

PaRC-mix training was compared against various mixup training methods including (a) without mixup (no-mixup), (b) input mixup (input-mix), wherein features computed at the input layer are linearly combined (Zhang *et al.*, 2018), (c) incoherent mixup (incoherent-mix), wherein only features computed along the backbone layers of the networks were modified (Jung *et al.*, 2019). All models were trained using identical training and validation data (Table 1) as used in prior study (Simeth *et al.*, 2023). Disaggregated analysis was performed to evaluate impact of tumor size, ADC, and lesion aggressiveness on tumor detection and segmentation accuracy. Finally, feature centric analysis was performed to assess the reason for differences in generalization performance by computing t-stochastic network embedding (t-SNE) clustering of penultimate layer features to visualize differences in the features extracted by the models. Figure 1 shows a schematic of our approach. To support reproducible analysis, all models with weights will be available through open-source GitHub repository upon manuscript acceptance.

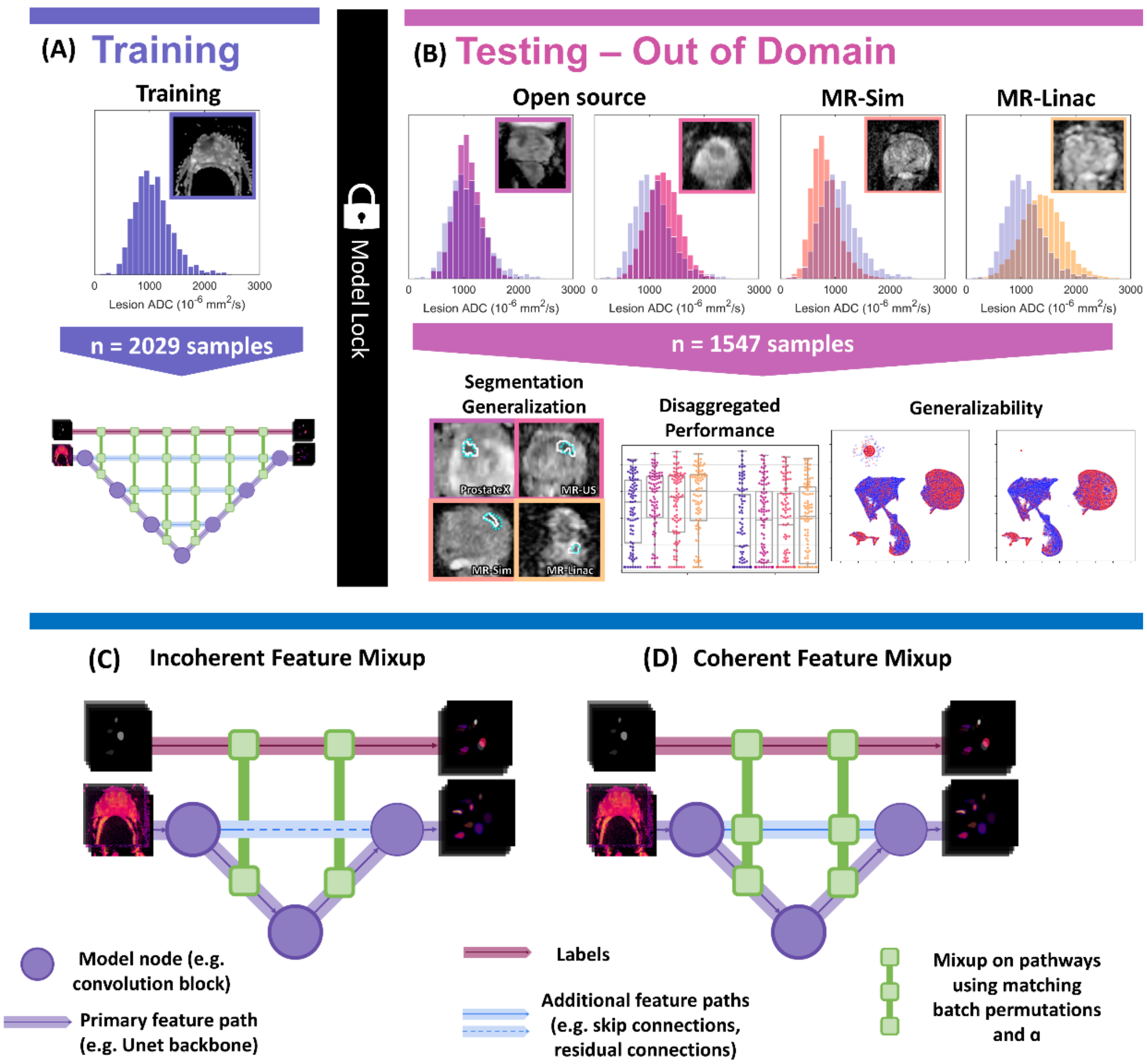

**Figure 1. (A) Model training and validation to select the best model was performed on an internal dataset containing 2029 examples. (B) All models were tested on four distinct out of domain datasets, totaling 1428 examples, provided to model only after training, to assess segmentation quality and lesion detection accuracy across datasets and lesion characteristics. (C,D) Multiple training strategies were implemented including coherent and incoherent feature mixup methods.**

## Related works

The focus of this paper is the segmentation of prostate cancer lesions from MRI with generalizability across multiple target domains. Several prior works exist for generalizable segmentation of the prostate gland (De Meerleer *et al.*, 2004; Lips *et al.*, 2011; Ippolito *et al.*, 2012; Aluwini *et al.*, 2013; Schild *et al.*, 2014; Dai *et al.*, 2020; Adams *et al.*, 2022; Zaridis *et al.*, 2024; Wu *et al.*, 2025; Zhang *et al.*, 2025) as well as detection of the prostate cancer from MRI (M. Y. Chen *et al.*, 2020; Dai *et al.*, 2020; Saha *et al.*, 2024), our discussion of related works pertains to those involving prostate cancer segmentation and medical image analysis methods focusing on domain generalization from single source training.

### Prostate cancer segmentation

Prior works have studied the segmentation of intra-prostatic prostate cancer (PCa) lesions (Jung *et al.*, 2019; Dai *et al.*, 2020; Y. Chen *et al.*, 2020; Matkovic *et al.*, 2021; Duran *et al.*, 2022; Eidex *et al.*, 2022; Alzate-Grisales *et al.*, 2023). Intra-prostatic lesions are hard to segment due to multiple reasons. First, high inter-rater delineation variability leads to inconsistencies in manual delineations used to train the segmentation models (Rischke *et al.*, 2013; Steenbergen *et al.*, 2015). Second, the lesions are hard to visualize on T2-weighted MRI and large differences in the appearance of lesions occurring in central versus peripheral gland, combined with relatively smaller size of lesions makes accurate localization and segmentation difficult. Probabilistic methods such the Probabilistic Unet and PHiSeg explicitly learn differences in expert segmentations and generate multiple segmentations (Kohl *et al.*, 2018; Baumgartner *et al.*, 2019). PHiSeg method has shown to be effective to visualize segmentation uncertainty for prostate zone segmentations. One key problem for lesions is the requirement of multiple rater segmentations to train such models, which are difficult to obtain in clinical practice. Another approach to overcome the difficulty of accurately localizing the lesions makes use of prostate gland segmentation to refine the localization of PCa lesions. Prior work by Saha et.al learned a probabilistic prior map of PCa locations within the prostate gland as part of training to reduce inaccurate detections (Saha, Hosseinzadeh and Huisman, 2021). Dai et al applied a mask RCNN, and reported results for single-source generalization, as well as with mixed domains. The observed substantially worse results when attempting to generalize to institutional data from open-source data than when they trained with mixed institutional and open source data, illustrating the difficulty of single-source domain generalization (Dai *et al.*, 2020).

### Single-source to multi-domain generalization

The goal of single-source domain generalization (SDG) is to train a model from a single data source such that it can generalize across data distribution shifts to unseen domains. In the context of medical imaging these distribution shifts are common when applying models trained at a single institution to other institutional data with differing scanners, protocols and populations. To enable generalizability across domains, approaches to SDG aim to train the model to use features that are invariant under domain shift. While novel loss functions (Panfilov *et al.*, 2019; Arslan, Guo and Li, 2024; Wen *et al.*, 2024) and model architectures (Saha, Hosseinzadeh and Huisman, 2021; Gao *et al.*, 2023; Simeth *et al.*, 2023) have been applied as means to aid in domain generalization, the main body of work has accomplished this through data diversification. Predicted domain shifts can be simulated (Singla *et al.*, 2021; Tirindelli *et al.*, 2021; Ouyang *et al.*, 2023) but, in general, variation will need to be stochastic (Xu *et al.*, 2020; Zhang *et al.*, 2020; Choi *et al.*, 2023; Aminbeidokhti *et al.*, 2024) or learned (Wang *et al.*, 2025; Weifeng, 2025). Variations can be introduced with standard augmentations (Zhang *et al.*, 2020; Tirindelli *et al.*, 2021; Aminbeidokhti *et al.*, 2024), augmentations in frequency (Yang and Soatto, 2020; Choi *et al.*, 2023; Li *et al.*, 2023; Arslan, Guo and Li, 2024; Qiang, 2024) or feature space (Zhou *et al.*, 2021; Kang *et al.*, 2023; Huang, Han and Dou, 2025; Tyler, 2025) , or even adversarial example training (Xu *et al.*, 2022; Chen *et al.*, 2023; Cheng, Gokhale and Yang, 2023; Gokhale *et al.*, 2023; Wang *et al.*, 2023). This class of approaches can enhance generalizability across the many data distribution shifts when generalizing to an unseen domain and can generally be used in concert with other approaches to SDG. However, most augmentation approaches will be sensitive to hyperparameter tuning, needing to find the correct level of perturbation to establish a realistic, or at least useful, training distribution. Mixup methods have provided a data agnostic approach that provides robust performance with minimal hyperparameter tuning.

### Mixup training strategies for network regularization

Mixup is a network training regularization approach that provides augmented examples by linearly combining separate training images and performing a matching combination for the image labels(Zhang *et al.*, 2018). Co-mixup is another approach that improves on input mixup methods by combining input image patches from all images available in the training batch to generate augmented images and labels. Co-mixup methods have been successfully used for natural image classification tasks (Kim *et al.*, 2021). Manifold-mixup applies the same mixup operation as Zhang et al (Zhang *et al.*, 2018) but in the learned feature space (Verma *et al.*, 2019). Our approach is related to and extends the manifold or feature mixup methods that linearly combine features extracted from a stochastically selected pair of input images. At each iteration manifold mixup augmentation randomly selects a single network layer from a list of eligible layers and performs feature mixup using that layer (Verma *et al.*, 2019). This results in a limitation that only features of one network layer are modified at each training iteration. Jung et.al introduced stochastic selection of the feature layer, with the possibility of application at multiple layers that occupied the network backbone (Jung *et al.*, 2019). However, this approach explicitly selected specific encoder or decoder or bottleneck layers to perform mixup training (Jung *et al.*, 2019), without any applying mixup coherently on both the network backbone and the parallel feature streams of the skip connections. Our approach applies mixup at multiple network locations on each training iteration while concurrently applying the operation at each parallel feature stream, maintaining coherence between all feature streams and labels.

## METHODS

### Deep learning architectures

Two convolutional neural networks that use dense and residual connections were analyzed, (1) a multiple resolution residual network (MRRN) (Jiang *et al.*, 2019) and (2) UNet++ (Zhou *et al.*, 2018). MRRN was originally developed for lung tumor segmentation from computed tomography scans and uses dense and residually connected feature streams to combine image features at various levels and image resolutions to handle tumors of varying sizes. We extended this model for segmenting prostate cancers from MRI (Simeth *et al.*, 2023). UNet++ uses nested subnetworks with dense skip connections in order to carry the features from multiple resolutions while spanning the encoder/decoder semantic gap along with deep supervision at the various feature layers to regularize feature extraction. Convolutional architectures were analyzed in order to leverage the local spatial hierarchy of the features for segmenting prostate cancers that always occur within the prostate gland.

As shown in Figure 2, MRRN implementation used the default network architecture and consisted of 3 encoder residual units followed by 4 encoder convolutional blocks and 3 decoder convolutional blocks followed by 3 decoder residual units, with 4 residual feature streams used for the MRRN. The UNet++ consisted of 4 subnetworks, with 9 convolution nodes in the main network backbone, along with 6 additional nodes forming the decoder elements for the 3 additional subnetworks, and dense skip connections forming 4 additional feature streams for carrying information from the encoders at different resolution levels. This network is illustrated in our prior work (Simeth *et al.*, 2026). Multiple mixup training methods were implemented into the network training. Mixup training does not alter the network architectures, and, following training, the default networks were used for testing.

### Feature mixup augmentation for training deep networks

We sought to increase the diversity of computed features by creating variations in features computed at the encoder, bottleneck, and decoder layers. We distinguished the feature mixup performed along the network backbone layers as incoherent mixup versus mixup augmentation performed for both backbone and intermediate feature streams used to feed features into the backbone layers as coherent mixup strategies. Incoherent mixup is “incoherent” because the intermediate feature streams do not undergo mixup and hence the computed feature representations do not correspond to feature variations generated at the backbone layers or label identities.

The coherent mixup strategy used in PaRC-mix performs a non-linear mixup by sequentially varying features at all the feature layers occurring in the backbone as well as intermediate layers from deep skip and residual feature streams to produce a rich representation of the computed features. Hence, this approach allows to extract a more comprehensive variation of the data in various feature resolution and semantic context levels.

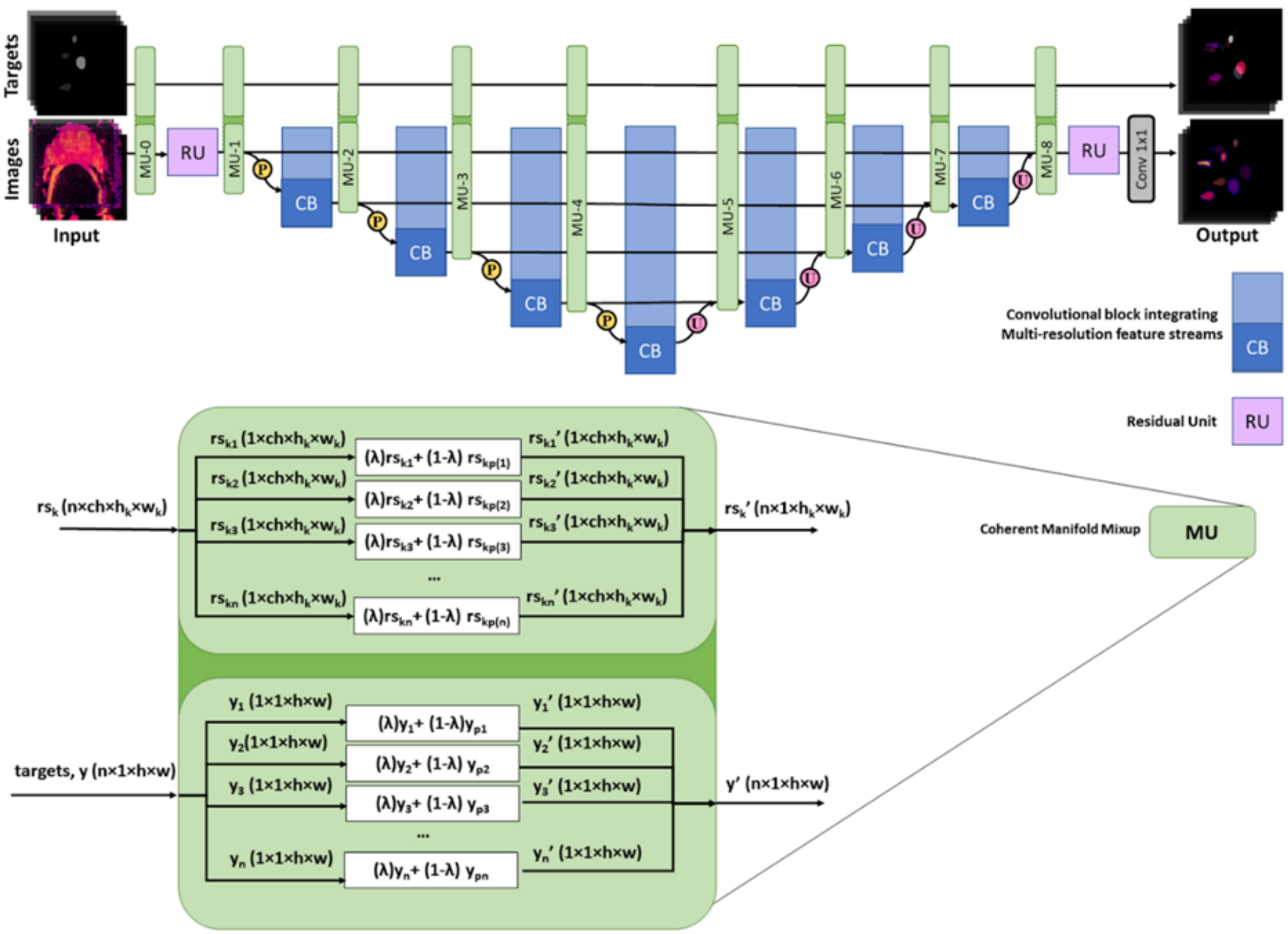

**Figure 2. The architecture for MRRN with PaRC-mix. PaRC-mix can include any combination of layers undergoing mixup. RU corresponds to the Residual Unit. For each feature stream $rs_k$ passing through a mixup layer the features in each sample of the batch are interpolated using the weighting parameter (drawn from the beta distribution. β(α, α)), with another random sample from the batch that is held constant for all concurrent interpolations for each feature stream and the targets for each sample ($y_i$).**

Given a candidate layer k in the network, we group all features from residual and skip connections that pass through that feature layer $X_k = (x_{k1}, x_{k2}, \dots x_{kn})$ along with the corresponding labels y. In Figure 2, the features in $X_k$ are the feature streams (lines and arrows) that pass through mixup block k (green box). We perform coherent mixup by paired interpolation of features and labels from two samples $(X_k, y_k)$ and $(X_k', y_k')$:

$$X_{mixed} = X * \lambda + X' * (1-\lambda)$$

$$y_{mixed} = y * \lambda + y' * (1-\lambda)$$

Where **λ** is the random mixup weight derived from the beta distribution. Instead of being performed on the grouping of all parallel features $X_k$, Incoherent-mixup operates only on a subset of the feature streams, most often the single feature stream of the network backbone, e.g., $X_{k_backbone}= (x_{kk})$. This implementation of manifold mixup results in incoherence between the omitted features and the rest of the network, including the target labels.

In our implementation we create the minibatch $(X'_k, y'_k)$ by randomly shuffling the rows of the batch $(X_k, y_k)$, resulting in a random pairing of the samples in the superbatch. Our mixup approach allows a flexible mixup strategy spanning multiple feature layers, creating a non-linear mixup of more than a single pair of batch elements.

We investigate four training methods for both the MRRN and UNet++ network which differ only in mixup strategy:

1. No-mixup, in which the network is trained with only standard augmentations such as image rotations, scaling, color jittering, etc.
2. Input-mix, wherein in addition to standard augmentation the input images and labels are subject to mixup.
3. Incoherent-mix, which also incorporates feature mixup along the feature layers occurring in the network backbone, and
4. PaRC-mix, which incorporates coherent feature mixup along all parallel paths or features layers in the network.

**Implementation**

All models were implemented in Pytorch (Paszke *et al.*, 2017) and optimized with Adam (Kingma and Ba, 2017) using a learning rate of 0.0002 for MRRN and 0.0001 for UNet++. Training began with a constant learning rate for the first 20 epochs, then linearly decaying towards zero at 600 epochs. The label smoothing hyperparameter $\epsilon$, batch size, and mixup hyperparameters were determined empirically by best segmentation performance defined as the model with the highest normed score on the validation dataset. Label smoothing was applied with $\boldsymbol{\epsilon}$=0.001. Batch size was set to eight. Mixup weights $\boldsymbol{\lambda}$ were drawn from the beta distribution $\boldsymbol{\beta}(1,1)$, which corresponds to a uniform distribution.

Feature mixup was implemented at the encoder, bottleneck, and decoder as well as input layers for MRRN method. Hyperparameter analysis was performed to identify the best combination of such features, which identified combining the input (MU-0) and bottleneck layers (MU-4 and MU-5) to provide the best performance (see Figure 2). Feature mixup analysis for UNet++ showed that mixing features from all blocks resulted in the best performance and all blocks were thus selected and used for feature mixup (see Figure 3). Note that implementation at multiple layers with a relatively flat beta distribution is only viable with certain loss functions. Soft Dice loss was used to ameliorate this concern, since it produces predictions near 0 or 1, creating predictions even for fairly attenuated labels. Conventional augmentation was also performed through left-right flips, scaling, rotation, and elastic deformation.

All segmentation models were trained with identical training datasets and optimized with a soft Dice loss. The model was trained and evaluated on cropped regions of 128×128×5. Test time lesion metrics were computed on the reassembled slices. Models were trained using all slices containing intra-prostatic lesions. In addition, up to 5 slices occurring above and beneath the lesions were included to provide negative examples. Early stopping was used to prevent overfitting. Manifold mixup was implemented by performing a matching mixup procedure to an arbitrary list of feature tensors along with the target. This allows matched mixup to be applied to each parallel feature stream concurrently.

**Table 1. Number of intermediate to high risk (IHR) lesions for each dataset (counting lesions contained in separate scans multiple times), and the corresponding lesion characteristics in terms of median volume and**

**ADC with interquartile range in brackets. Note that the images in the training and validation set used an endorectal coil (EC). Significant differences in lesion volume or ADC from the training set are marked with an *.**

| Dataset | Scanner | Number of IHR or PIRADS≥ 4 Lesions (GS ≥ 7) /Total | Median Lesion Volume (cc) | Median Lesion ADC ($\times10^{-3}$ $mm^2/s$) | Resolution ($mm^3$) |
|---|---|---|---|---|---|
| **Training/ Validation** | **Internal Diagnostic w/ EC, 3T GE** | **147/191** | **0.84 [0.39-1.3]** | **0.95 [0.87-1.11]** | **0.625×0.625×3 and 0.78×0.78×3.6** |
| **ProstateX** | **External Diagnostic, 3T Siemens** | **76/298** | **1.1 [0.51-2.6]** | **0.96 [0.88-1.04]** | **2×2×3 and 2×2×4** |
| **Prostate158** | **External Diagnostic, 3T Siemens** | **119/119** | **1.35 [0.67 – 2.8]*** | **1.58 [1.41-1.82]*** | **0.272×0.272×3 to 0.498×0.622×4.305** |
| **MR-US** | **External Diagnostic, 3T Siemens** | **492/871** | **0.45 [0.17-1.3]*** | **1.16 [1.04-1.33]*** | **1.625×1.625×3.6** |
| **MR-Sim** | **Internal 3T Philips** | **59/59** | **0.58 [0.26-1.8]** | **0.81 [0.69-0.94]*** | **1.125×1.125×4** |
| **MR- Linac** | **Internal 1.5T Philips** | **200/200** | **1.52 [0.99-2.4]*** | **1.43 [1.28-1.53]*** | **1.92×1.92×4** |

### Datasets characteristics used in training and testing

Six datasets were used. An institutional dataset consisting of diagnostic MRI was used for training and validating the models. Four different datasets consisting of two external datasets using diagnostic MRI, third dataset consisting of MR simulation institutional scans acquired prior to radiation treatment planning and the fourth testing dataset consisting of MRI acquired during radiation treatment in patients with prostate cancers who underwent MR guided adaptive hypofractionated radiation treatment in an MR-Linac were analyzed. Descriptions of the six datasets are provided below with detailed clinical scan characteristics summarized in table 1.

1. **Training and validation:** consisted of data acquired on diagnostic scanners (3T GE) from 151 patients with 191 biopsy proven prostate cancers. T2-weighted (T2w) and ADC maps are obtained and were delineated by an expert radiologist, using both T2w and ADC maps in consensus with two pathology research fellows with whole-mount step section pathological tumor maps from surgically excised prostate glands. Each scan used an endorectal coil to enhance image quality. This dataset is described in more detail in prior research (Fehr *et al.*, 2015; Simeth *et al.*, 2023). Prostate centered cropping of 128×128×5 regions resulted in 2029 samples for training of the models.

2. **ProstateX:** External diagnostic testing data acquired on diagnostic scanners (3T Siemens)

was derived from the ProstateX grand challenge dataset (Litjens *et al.*, 2017; Armato *et al.*, 2018). These retrospective prostate MR studies include ADC maps acquired on Magneton Trio and Skyra, two Siemens 3T MRI scanners. Publicly available manual segmentations from Cuocolo et al (Cuocolo *et al.*, 2021) were used for model testing.

**3. Prostate158:** External diagnostic testing data acquired on diagnostic scanners (3T Siemens) (Adams *et al.*, 2022). These retrospective prostate MR studies include ADC maps with contours for 119 cases with PCa- with PIRADS $\geq$ 4, and for two separate reviewers for a subset of 20 cases.

**4. MR-US:** External diagnostic testing data acquired on diagnostic scanners (3T Siemens) was derived from The Cancer Imaging Archive (TCIA) Prostate-MRI-US-Biopsy dataset. This 1,151 patient dataset included ADC scans for a subset of 837 patients. MRI targets were segmented and scored based on the suspiciousness of the lesion on a 5-point Likert scale similar to PIRADS. Lesions were contoured on corresponding T2w MR. Scans where the provided ADC to T2w alignment in 3DSlicer placed the lesion contour outside the prostate were excluded from analysis (Choyke *et al.*, 2016).

**5. MR-Sim:** Internal MR-Sim testing dataset (MR-Sim interrater data, 3T Philips) was composed of 59 MRI studies acquired prior to radiation therapy. Two radiologists (one with 15 years and the other with 5 years experience) performed independent evaluation and segmentation of seventeen different cases with intra-prostatic lesions in order to compute inter-rater variability in segmentation. Clinical lesions were designated on T2w MR with reference to ADC available.

**6. MR-Linac:** Internal MR-Linac testing dataset (MR-Linac treatment data, 1.5 Tesla Philips) consisted of 40 patients from dataset IV with an additional 5 MR-Linac scans each corresponding to 5 fractions of treatment. Each image included lesion designations and RT dose information. Clinical lesions were designated on T2w MR with reference to ADC available.

**Evaluation metrics**

Lesion detection accuracy was measured using precision and recall metrics, where lesions were considered to be detected if the model segmentations had a DSC overlap of at least 0.1 with respect to manual delineations as previously used in Saha et al (Saha, Hosseinzadeh and Huisman, 2021), Mckinney et al (McKinney *et al.*, 2020), and Simeth et al (Simeth *et al.*, 2023). Manual segmentations with volumes under 0.15 $cm^3$ were not evaluated.

Segmentation accuracy was measured using multiple metrics including DSC, 95$^{th}$ percentile Hausdorff distance (HD95), and mean surface distance (MSD). Only lesions that were detected were evaluated for segmentation accuracy. In order to interpret results when using multiple metrics and in the context of inter-rater variability in generating the manual delineations, we computed a composite metric based on the method seen by Yang et al (Yang *et al.*, 2018) that explicitly uses interrater variability in computing the metric. Inter-rater variability in segmentation was obtained from a subset of 17 cases from the MR-Sim datasets that contained two rater delineations. Inter-rater variability is used to scale accuracy of the method with respect to the manual ground truth and is computed for each metric as,

$$Normed\ Metric\ Score\ =\ max\left(50 + \frac{T - R}{P - R} \times 50{,}0\right)$$

Where P is the perfect metric score (1 for DSC, 0 for HD95 and MSD), R is the reference measure for that metric, and T is the test metric. The final combined normed score including all metrics is obtained by the average of normed scores from each metric.

Performance gap metric was computed by differencing average normed scores in the disaggregated analysis to measure performance differences of individual methods with respect to lesions with: (a) highly aggressive cancers defined as having a Gleason score (GS) greater than or equal to 4+3 from the least aggressive cancers defined as having a GS of 3+3 and (b) cancers occurring in the transition zone (TZ) from cancers occurring in the peripheral zone (PZ).

**Statistical analysis**

Statistical comparisons were conducted with paired, two-sided Wilcoxon signed-rank tests with the normed score when comparing different methods at 95% significance level. Unpaired, two-sided Wilcoxon signed-rank tests using normed scores were used to measure differences of individual methods in the disaggregated analysis involving prostate lesion Gleason scores and lesion locations (PZ versus TZ). Statistical analyses were performed using Matlab version R2020b. $p < 0.05$ was considered statistically significant. Adjustment for multiple comparisons used the Bonferroni method. Dependency of DSC metric on continuous metrics, namely, tumor volume and ADC was computed using Spearman rank correlation coefficient, and categorical metrics, namely, tumor Gleason scores (3+3 versus 4+3 or higher) and location (TZ versus PZ).

**Experiments:**

**Generalizability of model segmentation accuracy by scanner and image type:** The generalizability of model performance across datasets was evaluated by evaluating segmentation accuracy (in terms of DSC, HD95 and MSD) for (1) ProstateX, 3T Siemens data (2) MR-US, 3T Siemens data (3) MR-Sim, 3T Philips data (4) MR-Linac, 1.5T Philips data

**Versatility of implementing PaRC-mix:** We implemented PaRC-mix into two different models MRRN and UNet++. Additional mixup training methods were implemented into these methods including (1) no-mixup (2) input-mix (3) incoherent-mix applied to the encoding backbone.

**Disaggregated segmentation performance:** Differences in model performance under different relevant clinical conditions such as cancer aggressiveness (GS ≧ 4+3 versus GS = 3 + 3) as well as lesion location (peripheral versus transition zone) was analyzed to understand robustness of the models.

**Finetuning of model on clinical dataset:** We evaluated the potential for further improving PaRC-mix trained models for clinical datasets by finetuning the UNet++ model on the MR-Linac dataset with 5-fold cross validation. This experiment was performed to assess the best possible accuracy that could be reached for clinical use. Training in each fold was performed on 32 patients (roughly 160 fractions and 900 slices) and validated on 8 patients (roughly 40 fractions and 200 slices). Training was performed identically to the original model, except with iterative unfreezing of the network layers. Initially only the output convolutions were unfrozen, with nodes (0,3), (0,4) and (1,3) unfrozen after 2 epochs, (0,2), (1,2) and (2,2) unfrozen after 10 epochs, (0,1), (1,1), (2,1) and (3,1) unfrozen after 20 epochs, and all nodes unfrozen after 40 epochs.

**Reasons why models generalize differently:** The separability of the extracted features corresponding to tumor or non-tumor background voxels was compared across methods and datasets using t-SNE clustering. The compared features for each method were extracted from the penultimate layer. t-SNE analysis was performed in python using openTSNE with the cosine distance metric, a perplexity of 25, and FFT descent for 500 iterations after PCA initialization. Features were selected from the penultimate layer of each network. For each dataset, 500 voxels from each clinically significant lesion were aggregated with the same number of non-tumor voxels randomly selected from near the tumor. The area near the tumor was defined by the non-intersecting voxels from the dilation of the tumor volume by 5 and 10 voxels in-plane. t-SNE was performed on the internal validation data, and then partial embedding was used for a further 250 iterations to embed the points from each additional dataset into that same feature space.

**Data availability**

The open-source data that support the findings of this study are available on TCIA. The non-TCIA institutional datasets used in this study are currently restricted from public release due to data privacy laws and institutional review board policies.

**Code availability**

The underlying code for this study will be made available on GitHub at publication. Implementation of the MRRN model is available through GitHub (https://github.com/The-Veeraraghavan-Lab/MedPhys_PCA_Segmentation_MRRN_DS_2023).

## RESULTS

### PaRC-mix training resulted in higher segmentation accuracy across all testing datasets

Table 2 shows the accuracies produced by various methods together with accuracy reduction (↓) or accuracy gain (↑) of individual methods with respect to PaRC-mix. Pooled analysis combining all testing datasets showed that MRRN trained using PaRC-mix was more accurate than no-mixup training ($p<0.001$ for normed score, recall, and precision). When compared to input-mix, PaRC-mix produced similar segmentation ($p = 1$) but higher recall ($p = 0.037$) and precision ($p<0.001$), indicating more detections with fewer false detections, an important requirement to reduce unnecessary edits by clinicians in adaptive workflow. On the other hand, when compared to incoherent mix, PaRC-mix produced more accurate PCa segmentations with higher recall ($p = 0.016$), normed score ($p < 0.001$), but a similar precision ($p=0.11$).

UNet++ trained with PaRC-mix was significantly more accurate than no-mixup ($p<0.001$ for normed score, recall, and precision) and input-mix ($p<0.001$ for normed score and recall; $p=0.002$ for precision). Like MRRN, PaRC-mix applied to UNet++ produced more accurate segmentations than incoherent mix indicated by higher normed score ($p<0.001$) and recall ($p<0.001$) with similar precision ($p=0.64$). Taken together, these results indicate that PaRC-mix training improves the accuracy of detecting and segmenting PCa lesions compared to other training methods. Detailed results with all analyzed metrics are in Extended data Figure E1.

Segmentations for representative examples from multiple testing sets produced by the methods (cyan dotted line) and manual delineations (white solid line) are shown in Figure 3. As shown, models trained with any mixup approach were more accurate than no-mixup training. PaRC-mix delineations closely followed the diffusion restricted intra-prostate regions, which typically harbor aggressive lesions (Salami *et al.*, 2017). Prostate158 included multiple reader delineations for a subset of images. Extended data Figure E2 shows several example volumes (axial, sagittal, and transverse orientation) depicting the segmentation from PaRC-mix MRRN along with both readers. Reported metrics refer to reader 1.

In order to assess robustness to scanner variations, PCa lesion segmentation results for same patients scanned on MR-sim for treatment planning simulation and on the MR-Linac on the first treatment fraction prior to commencing radiation treatment are shown in columns 6 and 7. As shown, in all cases, the models resulted in an under segmentation compared to clinical delineation on the MR-Linac cases (column 7) as AI segmentations were confined to the diffusion restricted region. Of note, the models were trained using radiologists' segmentations whereas, the MR-Linac datasets were delineated by radiation oncologists. Clinically, lesion margins are often extended to either account for microscopic disease or for motion of the lesions during treatment. Hence, DL model evaluation in this setting included an additional challenge of label shift.

For a subset of cases treated on the MR-Linac, radiation dose information was available for each treatment fraction. Hence, we compared the dosimetric accuracy of the MRRN trained using PaRC-mix compared to the same metric when computed using clinical delineations. The SALT-lomax Conformity index (CI)(Lomax and Scheib, 2003) was defined as

$$CI = TV_{RI}/TV$$

where $TV_{RI}$ is the tumor volume receiving at least the target dose (45Gy for boosted lesion) and TV is the volume of the target – the DIL in this case. AI conformity index had a high Spearman rank correlation coefficient of 0.875 ($p < 0.001$) with respect to the clinical conformity index, indicating good correlation between AI and clinical delineation derived dose metrics.

**Table 2. The median of normed segmentation scores comparing all lesions of GS ≥ 7 detected by at least one model (or PIRADSv2 ≥ 4 for MR-US) are shown in (A). Models underperforming PaRC-mix are indicated with blue down arrow and those over-performing PaRC-mix are indicated by red up arrow. (B) shows precision and (C) shows recall metrics for all analyzed models and datasets. Models with significantly different scores compared to PaRC-mix are indicated by * for p < 0.05, ** for p < 0.01, *** for p < 0.001 after adjusting for multiple comparisons using the Bonferroni method.**

| (a) | | ProstateX, 3T Siemens | Prostate158 | MR-US, 3T Siemens | MR-Sim, 3T Philips | MR-Linac, 1.5T Philips | All |
|---|---|---|---|---|---|---|---|
| | | Normed Score (diff) | Normed Score (diff) | Normed Score (diff) | Normed Score (diff) | Normed Score (diff) | Normed Score (diff) |
| MRRN | No-mixup | 56.8 (↓6.9**) | 35.5 (↓3.5) | 23.4 (↓16.5***) | 33.1 (↓8.6) | 15.2 (↓7.8) | 25.9 (↓13.3***) |
| | Input-mix | 62.4 (↓1.3) | 41.4 (↑2.1) | 42.2 (↑2.3) | 39.7 (↓2.0) | 15.2 (↓7.8) | 40.4 (↑1.8) |
| | Incoherent-mix | 58.7 (↓5.0) | 25.2 (↓14.1**) | 33.9 (↓6.0**) | 28.4 (↓13.3) | 13.3 (↓9.7***) | 31.2 (↓7.4***) |
| | PaRC-mix | 63.7 | 39.3 | 39.9 | 41.7 | 23.0 | 38.6 |
| UNet++ | No-mixup | 57.3 (↓4.1*) | 34.5(↓1.8*) | 29.7 (↓14.4***) | 36.2 (↓10.5) | 18.7 (↓2.2) | 30.6(↓8.8***) |
| | Input-mix | 57.5 (↓3.9) | 37.6(↑1.3) | 39.8 (↓4.3**) | 38.5 (↓8.2) | 19.5 (↓1.4) | 36.6 (↓2.8***) |
| | Incoherent-mix | 67 (↑5.6) | 36.7(↑0.4) | 39.6 (↓4.5***) | 21.1 (↓26.6) | 16.9 (↓4.0**) | 37.7 (↓1.7***) |
| | PaRC-mix | 61.4 | 36.3 | 44.1 | 46.7 | 20.9 | 39.4 |

| (b) | | ProstateX, 3T Siemens | Prostate158 | MR-US, 3T Siemens | MR-Sim, 3T Philips | MR-Linac, 1.5T Philips | All |
|---|---|---|---|---|---|---|---|
| | | Precision | Precision | Precision | Precision | Precision | Precision |
| MRRN | No-mixup | 0.22*** | 0.26 | 0.22*** | 0.22 | 0.16 | 0.21*** |
| | Input-mix | 0.26 | 0.22 | 0.25* | 0.21 | 0.14 | 0.22* |
| | Incoherent-mix | 0.33 | 0.3 | 0.3 | 0.29* | 0.19 | 0.28* |
| | PaRC-mix | 0.31 | 0.28 | 0.28 | 0.25 | 0.18 | 0.26 |
| UNet++ | No-mixup | 0.21* | 0.2 | 0.20*** | 0.18 | 0.16 | 0.19*** |
| | Input-mix | 0.24 | 0.2 | 0.22 | 0.18 | 0.14 | 0.21** |
| | Incoherent-mix | 0.29 | 0.27 | 0.26 | 0.2 | 0.17 | 0.24 |
| | PaRC-mix | 0.27 | 0.23 | 0.25 | 0.21 | 0.17 | 0.23 |

| (c) | | ProstateX, 3T Siemens | Prostate158 | MR-US, 3T Siemens | MR-Sim, 3T Philips | MR-Linac, 1.5T Philips | All |
|---|---|---|---|---|---|---|---|
| | | Recall | Recall | Recall | Recall | Recall | Recall |
| MRRN | No-mixup | 0.83** | 0.84 | 0.71*** | 0.73 | 0.67 | 0.76*** |
| | Input-mix | 0.97 | 0.9 | 0.89 | 0.73 | 0.75 | 0.89* |
| | Incoherent-mix | 0.91 | 0.83 | 0.8 | 0.73 | 0.68 | 0.82* |
| | PaRC-mix | 0.99 | 0.85 | 0.85 | 0.8 | 0.69 | 0.86 |
| UNet++ | No-mixup | 0.87* | 0.81 | 0.81*** | 0.69 | 0.78 | 0.83*** |
| | Input-mix | 0.96 | 0.9 | 0.85*** | 0.75 | 0.69** | 0.85*** |
| | Incoherent-mix | 0.99 | 0.93 | 0.85*** | 0.68 | 0.72 | 0.86*** |
| | PaRC-mix | 1 | 0.91 | 0.93 | 0.8 | 0.78 | 0.92 |

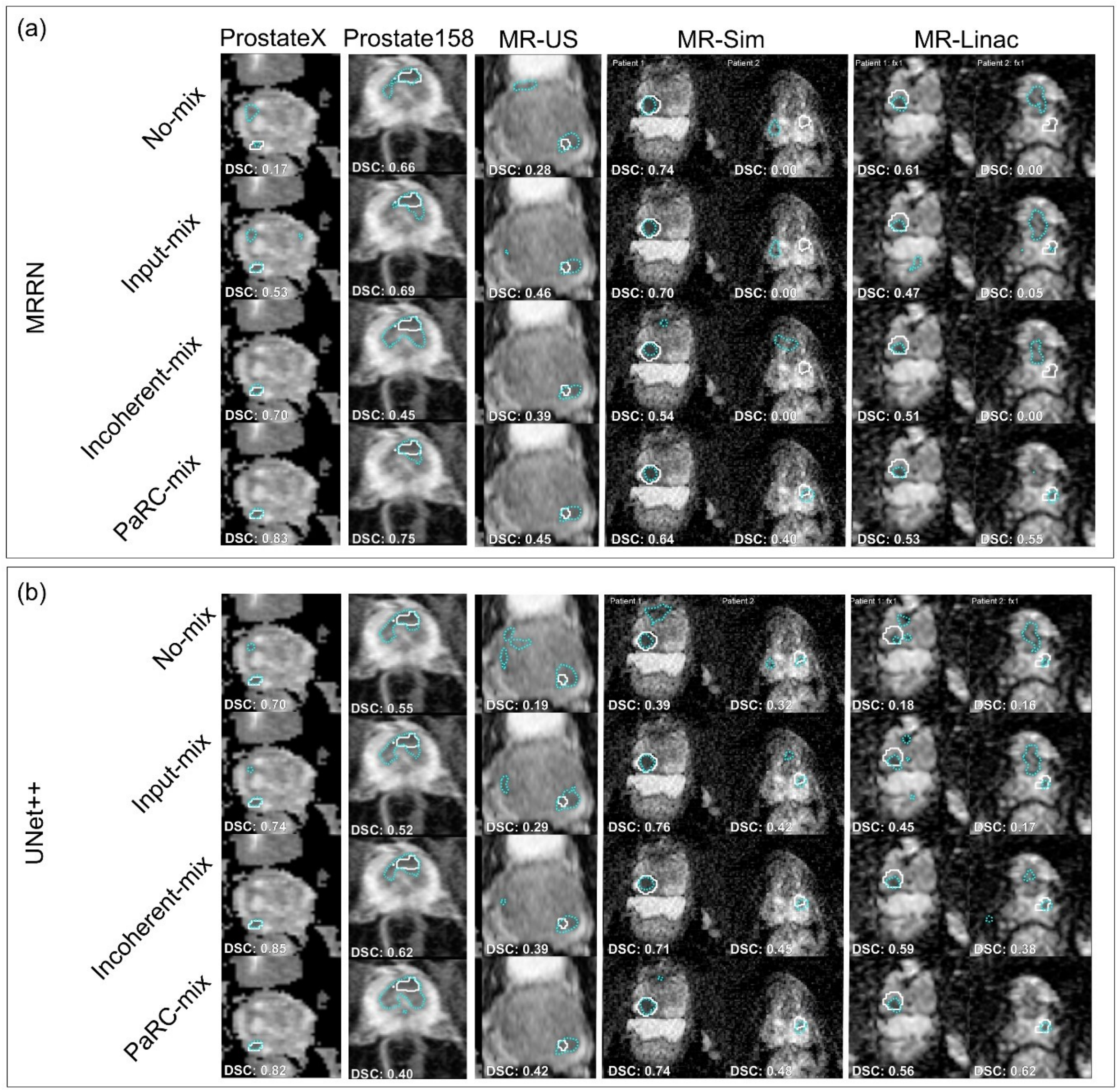

**Figure 3. Representative examples from analyzed MR scanners used in the testing datasets showing segmentation results produced by various training strategies applied to MRRN (A) and UNet++ in (B). Manual delineation is depicted using solid "white" contour while model-generated segmentation is shown in dotted "cyan" color. The two columns of MR-Sim and MR-Linac show matching patients from MR-Sim and the first fraction of treatment on the MR-Linac.**

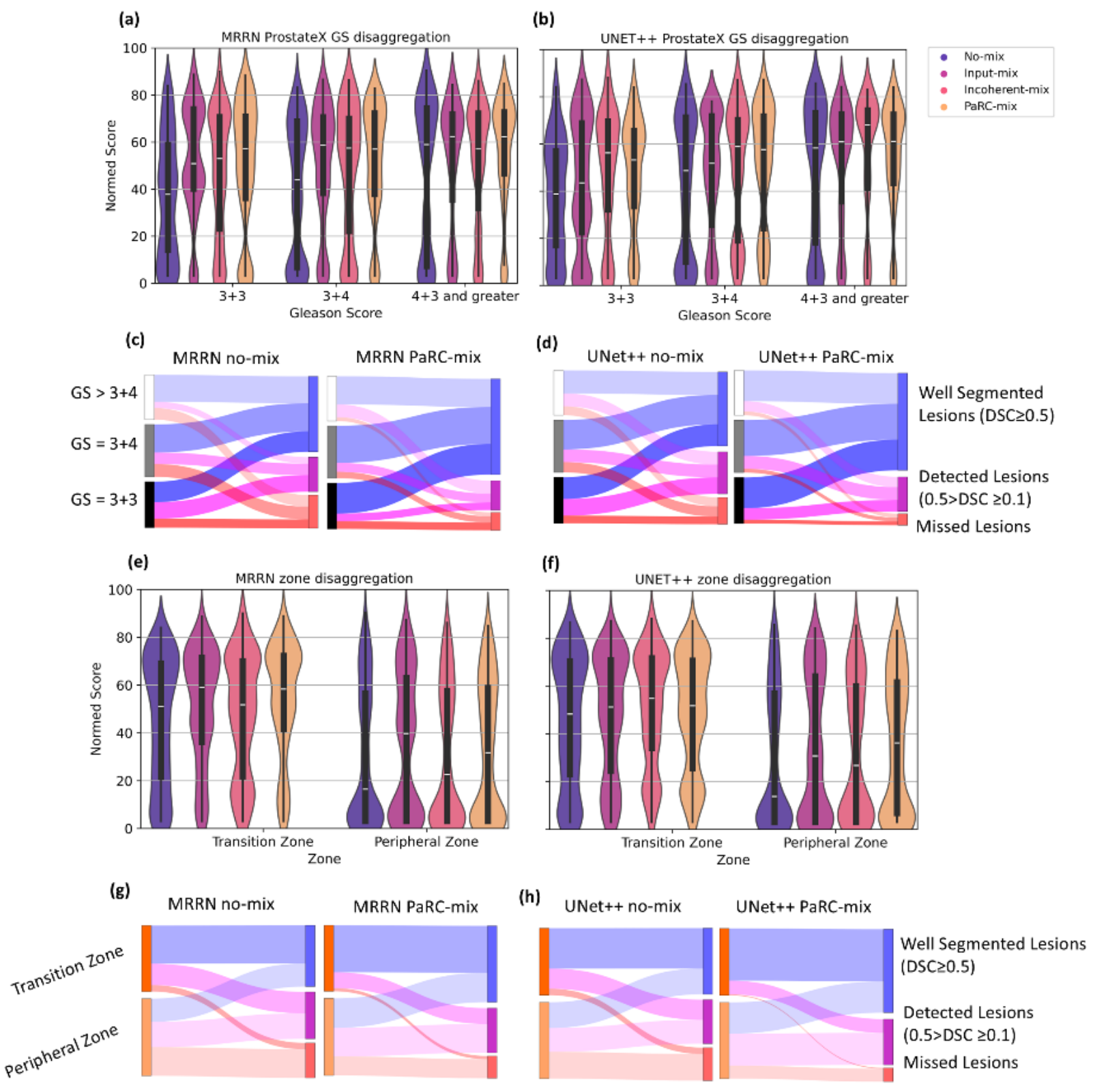

**Figure 4. Disaggregated analysis depicts the normed score segmentation accuracy for the analyzed methods with respect to Gleason scores of the segmented lesions in (A-B) and with respect to lesion location categorized as peripheral zone or PZ and transition zone or TZ (E-F). Sankey plots show the lesion breakdown for MRRN and UNet++ methods trained with PaRC-mix and no-mixup training in terms of lesion quality defined as well segmented (DSC ≥ 0.5), detected but poorly segmented (0.5>DSC≥0.1) and not detected (DSC<0.1) lesions. This is shown for lesions grouped by Gleason score (C-D) and location in the prostate (G-H).**

## Disaggregated analysis of models' segmentation accuracies with varying lesion characteristics

Lesion level GS were available for ProstateX; 35 lesions had GS 4+ 3 or higher, 41 had GS 3+4, and 36 had GS 3+3. Disaggregated analysis found PaRC-mix training of MRRN resulted in higher accuracy for both less aggressive (GS 3+3 tumors) and highly aggressive tumors (GS 4+3 or higher) compared to other methods (Figure 4 A-B). For UNet++, both PaRC-mix and incoherent-mix training achieved higher median accuracy than no-mixup or input-mix.

Performance gap was computed as the difference in the median normed score between highly aggressive (GS 4+3 or higher) and less aggressive (GS 3+3) lesions for the individual methods. Models trained with no-mixup resulted in the largest performance gap (MRRN of 21.1 and UNet++ of 19.5), indicating that these models were not robust to variations in lesion aggressiveness. Mixup training of MRRN reduced the performance gap, with input-mix achieving a gap of 11.6 versus 4.1 for incoherent-mix and 5.2 for PaRC-mix. A similar trend was seen for UNet++ with a performance gap of 17.4 for input-mix, 11.8 for incoherent-mix, and 7.9 for PaRC-mix.

Segmentation quality measured as well-segmented (DSC ≥0.5) versus detected (0.5>DSC≥0.1) and not detected (DSC<0.1) lesions relative to GS are depicted using Sankey plots (see Figure 4 C and D) for the different methods. PaRC-mix training increased the number of well-segmented lesions and reduced the number of undetected lesions across all groups for both MRRN and UNet++.

Lesion level zone designations were available for a subset of ProstateX, MR-Sim, and MR-Linac datasets. 83 lesions were designated in the peripheral zone (PZ) and 97 lesions in the transition zone (TZ). Lesions occurring in the anterior stroma were grouped with the TZ lesions. Lesions in the TZ were segmented with significantly higher accuracy compared to those occurring in the PZ irrespective of the training approach for both MRRN ($p < 0.001$ for all) and UNet++ (no-mixup $p<0.001$, input-mix $p=0.006$, incoherent-mix $p<0.001$, PaRC-mix $p = 0.003$) as shown in Figure 4 E-F. Performance gap analysis comparing normed scores for lesions occurring in TZ and PZ showed the largest gap for no-mixup training (33.3 for MRRN, 32.5 for UNet++). Performance gap reduced with mixup training for MRRN with 21.5 for input-mix, 20.8 for incoherent-mix, and 19.5 for PaRC-mix. In the case of UNet++, the performance gap decreased to 19.8 for input-mix, 29.3 for incoherent-mix, and 15.8 for PaRC-mix.

As shown in the Sankey plots (see Figure 4 G and H), PaRC-mix training applied to MRRN and UNet++ improved the proportion of well segmented lesions occurring in both the PZ and TZ locations. PaRC-mix training for both methods decreased the number of undetected and poorly segmented lesions in the TZ and undetected lesions in the PZ. Sankey plots for additional models are in Extended data Figure E3.

Finally, PaRC-mix DSC was moderately correlated with lesion ADC for both MRRN ($\rho$=-0.47) and UNet++ ($\rho$= -0.45). The correlation with volume was small for both MRRN ($\rho$=-0.04) and UNet++ ($\rho$= -0.09) (see Extended data Figure E4). These results indicate that mixup training especially using PaRC-mix as well as incoherent mix improved the robustness of the models to variations in lesion aggressiveness as well as spatial location within the prostate gland.

Moreover, UNet++ produced consistently higher recall for all external datasets compared to the MRRN model trained using PaRC-mix as seen in Table 2(c).

**Why does PaRC-mix generalize for multi-domain PCa segmentation?**

Reasons for performance differences between models were analyzed by visualizing the separation of features extracted within tumors and adjacent normal tissue using t-distributed Stochastic Neighbor Embedding (t-SNE) (Hinton and Roweis, 2002) of clinically significant lesions. Features from the penultimate network layer before the output were extracted. In each test example, up to 500 voxels were selected from within the segmented tumor and from normal tissue within an area defined by a 10 voxel in-plane dilation of the tumor, with the region defined by a 5 voxel in-plane dilation removed (forming a ring at a slight distance from the manually defined tumor). Similar embedding space was achieved for comparable visualization across all datasets using parameterized t-SNE embedding from the validation set of the in-distribution (or training) dataset and then applied on all the testing datasets.

Figure 5 shows t-SNE clustering results produced using the segmentations generated using the various analyzed methods for the various OOD as well as the ID datasets. Features corresponding to tumor and non-tumor classes were best separated for the ID data for all training strategies and models. Separation between the two classes becomes less pronounced for each OOD dataset, particularly MR-Linac data with no-mixup training. PaRC-mix training applied to MRRN and UNet++ resulted in the best separation for analyzed datasets compared to the other training methods even in the case of MR-Linac data. Finally, incoherent-mix training of the UNet++ did not extract features whose embeddings generalized to separate lesions from normal voxels in any of the analyzed testing datasets.

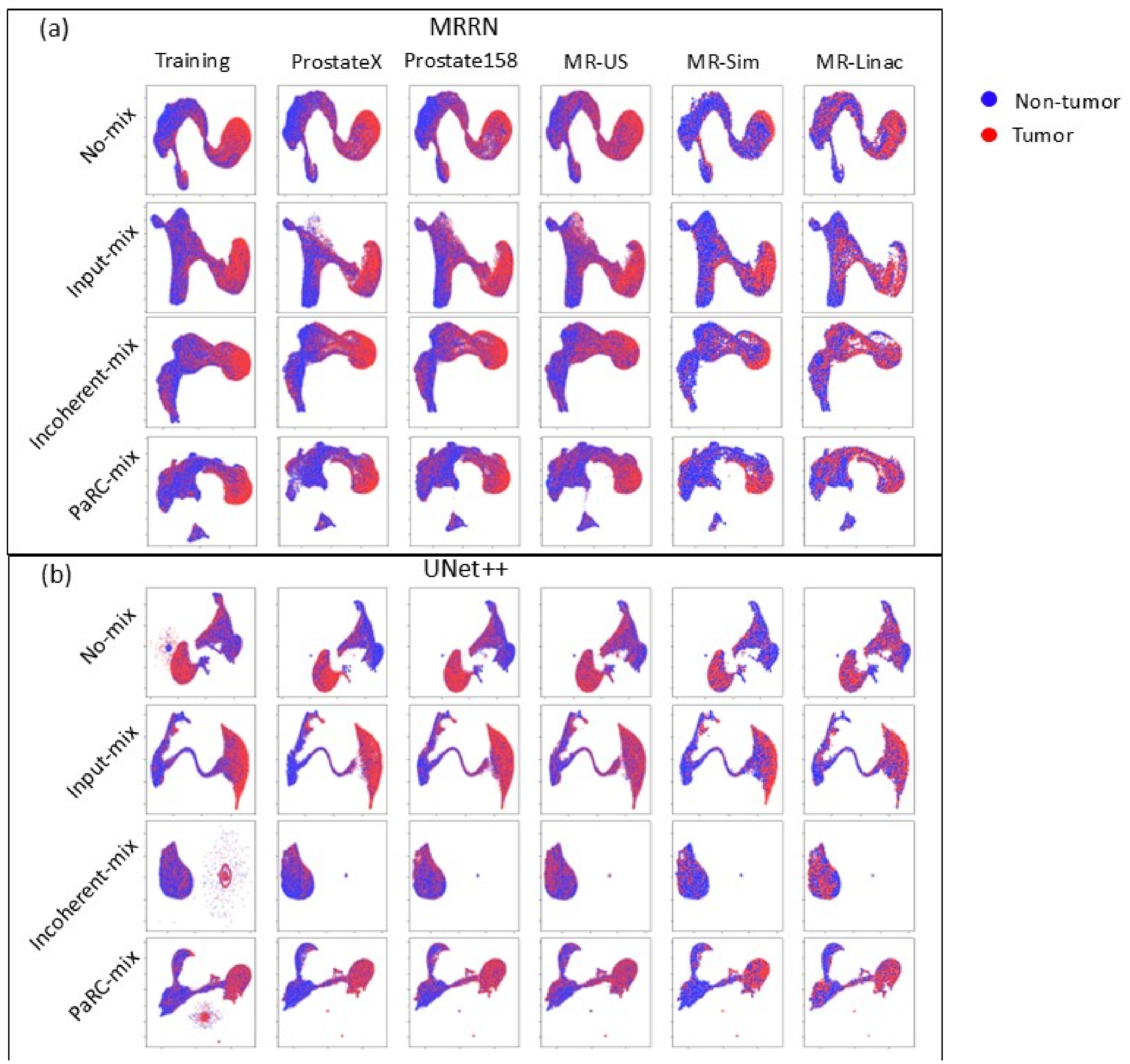

**Figure 5. t-SNE clustering of penultimate layer features' embeddings computed from models trained with no-mixup, input-mix, incoherent-mix, and PaRC-mix training strategies. Datapoints within the clinically segmented PCa (positives) are shown in red, normal tissue or negatives in blue. To aid visualization the training, ProstateX, Prostate158 and MR-US datasets use an opacity of 0.25 for displayed data points, while the smaller MR-Sim and MR-Linac datasets use an opacity of 0.8.**

### Fine-tuning further enhanced accuracy for clinical cases on the MR-Linac dataset

Finetuning of the UNet++ model via a 5-fold cross validated division of the 40 MR-Linac patients resulted in a significant increase in normed score as compared to the fine-tuned model (16.7 (8.6-39.1) without finetuning vs. 25.6 (9.5-51.7) with finetuning, $p<0.001$). This corresponded to a significant increase in DSC from 0.33 (0.18-0.43) to 0.38 (0.18-0.55), $p<0.001$, and decrease in HD95 from 7.76 mm (5.73 mm -11.84 mm) to 6.41 mm (4.75 mm - 9.82 mm), $p<0.01$. This was accompanied by a statistically significant drop in false positives, with precision increasing from 0.17 to 0.25 ($p<0.001$), but recall did not change significantly (0.78 vs 0.81, $p = 0.36$).

## DISCUSSION

We developed a novel training strategy that performs feature-level mixup using multiple network layers to increase the diversity of extracted features and thus improve multi-domain generalization even when the network is trained with single domain data. Our analysis of 4 different testing datasets with PCa lesions at various stages of diagnosis and treatment, lesion location, and aggressiveness showed that PaRC-mix approach overall improved the segmentation accuracy and detection ability compared to standard training methods. We demonstrated that our training approach is applicable to two different architectures. Our analysis showed that features extracted by the networks trained using PaRC-mix differed from those extracted by the same networks trained without mixup training contributing to larger variations in segmentation performance across datasets with different MR acquisitions. Our analysis also showed that networks trained using PaRC-mix better separated tumors from normal tissue voxels across all the various analyzed datasets, which contributes to its improved generalization capability. Furthermore, the PaRC-mix trained models resulted in more accurate PCa lesion delineations with higher precision and recall compared to models that didn't use mixup training across the different testing sets indicating robust generalization to different data distributions, cancer aggressiveness, and lesion locations. Evaluation with clinically relevant metrics such as conformity index showed strong correlation, indicating that PaRC-mix trained models could be translated for multi-institutional research and clinical studies.

In general, every form of mixup was better than when not using any mixup augmentation in training. Disaggregated analysis showed that PaRC-mix training resulted in higher accuracy even for less aggressive cancers. Furthermore, the accuracy gain and reduction of missed detections improved for lesions located in both transition and peripheral zones when using the PaRC-mix training strategy. This trend underscores the importance of feature-level augmentation in enhancing model generalization, surpassing the benefits of conventional augmentation techniques such as rotation, translation, and intensity jittering.

To our knowledge, this is the first PCa lesion segmentation approach that evaluated the generalization capability of a single source training method across a reasonably wide range of scanners and for patients scanned at different stages of disease and treatment, including diagnosis, planning simulation, and during hypofractionated radiation treatment. Importantly, our work also studied why generalization occurs with different training approaches. Prior works typically used data acquired from a single institution(M. Y. Chen *et al.*, 2020; Eidex *et al.*, 2022). A few prior works used data from different institutions for evaluation but were acquired with the same scanner manufacturer(Saha, Hosseinzadeh and Huisman, 2021; Duran *et al.*, 2022; Kohl *et al.*, no date). In our prior work, we evaluated the standard MRRN and multiple other methods on the ProstateX and a subset of the MR-Sim scans (Simeth *et al.*, 2023). In general, prior works, including ours, studied the capability of the models applied to diagnostic quality MRI as opposed to images acquired for treatment simulation and importantly during the course of radiation treatment. One prior work studied the application of deep learning methods for segmenting organs and the whole prostate gland used as the clinical target volume for radiation from 0.35 Tesla during treatment MR-Linac scans, using a patient specific transfer learning approach(Kawula *et al.*, 2023). Our approach, once trained on the 3.0 Tesla diagnostic quality GE scans using endorectal coil, was locked for testing on diagnostic and during treatment scans that did not include endorectal coil and used a variety of MR imaging acquisitions. Of note, prior works typically focused on segmenting the whole prostate gland, easier to segment than the intra-prostatic lesions due to the inherent difficulty in detecting the lesions and ensuing variability even between clinicians in delineating such lesions(Dai *et al.*, 2020; Liechti *et al.*, 2020; M. Y. Chen *et al.*, 2020). Furthermore, segmenting whole prostate gland is insufficient for dose escalation therapies that boost the dose to the dominant lesions, shown to be effective in reducing biological recurrence and increasing overall survival by clinical trials such as the multi-institutional FLAME trial.(Kerkmeijer *et al.*, 2021; Cock *et al.*, 2023)

Prior methods for domain invariant segmentation often used geometric priors, which are applicable to organs but not tumors due to their highly variable appearance and shapes. Adversarial perturbation of image intensities(Volpi *et al.*, 2018) and image style to alter underlying first and second order texture statistics (C. Chen *et al.*, 2020) can increase data diversity. However, such methods only vary the intensity statistics of the images themselves, which may not lead to sufficient diversity in the extracted features. Alternative methods such as feature-level augmentation implemented by linearly mixing input features from different domains (Zhou *et al.*, 2021) build on the input mixup methods (Verma *et al.*, 2019). Previously, Jung et al (Jung *et al.*, 2019) demonstrated enhanced accuracy by implementing feature level mixup separately for input, encoder, decoder, bottleneck and skip level layers of a UNet. However, this approach did not implement any measures to ensure coherence between manifold mixup in the network backbone and the long skip connections. Other feature-level mixup approaches employed feature mixup at each skip connection and output of the bottleneck layer (Zhu *et al.*, 2020) or randomly manipulated the feature layers used for mixup augmentation (Bdair *et al.*, 2022) implemented into the UNet in the context of semi-supervised learning. PaRC-mix improves over prior methods by performing feature level mixup coherently into encoder, decoder, residual and skip features within a single network. Our results show that PaRC-mix that uses coherent mixup of features for all paths results in better generalization compared to the incoherent mixup approach when considering both networks.

As a limitation, our results showed a drop in performance when the models were directly applied to the MR-Linac datasets, because this dataset represented challenges from domain shift as well as label shift due to differing norms for delineating lesions for radiation treatments relative to diagnosis by radiologists. We also found that this issue can be remedied through fine-tuning and for radiation treatments, it is possible to do patient-specific fine-tuning as needed to achieve highly accurate segmentation in consequent treatment fractions, reducing delineation requirements for radiation oncologists. This work focused on assessing the geometric and dosimetric accuracy of the AI generated segmentations compared to clinical delineations, but not in assessing if the AI segmentations further reduced clinical variability. Proper and detailed assessment of the same requires a larger and focused study and is planned future work. Our model was limited to using only the ADC instead of combining T2-weighted MRI to reduce impact of potential variations from T2-weighted imaging and limited accuracy improvement for combining the two modalities seen in our preliminary evaluation. Nevertheless, our results show reliable performance on multiple analyzed datasets with varied imaging acquisitions.

## CONCLUSION

We developed a novel feature-level mixup training approach that increases the reliability of models across widely different datasets. Our analysis showed that improved accuracy with our training approach extended to two different deep learning based convolutional networks. Furthermore, our methods resulted in robustly accurate segmentation irrespective of lesion aggressiveness and location, indicating that this is a potentially effective strategy for segmenting prostate cancer lesions for MR guided radiation treatments.

## ACKNOWLEDGMENTS

This work is supported in part by NIH grant P30 CA008748 and NIBIB R01EB032825. The authors thank the MSKCC imaging and radiation sciences (IMRAS) seed grant. This manuscript is the result of funding in whole or in part by the National Institutes of Health (NIH). It is subject to the NIH Public Access Policy. Through acceptance of this federal funding, NIH has been given a right to make this manuscript publicly available in PubMed Central upon the Official Date of Publication, as defined by NIH.

### Declaration of generative AI and AI-assisted technologies in the manuscript preparation process

During the preparation of this work the authors used ChatGPT and Claude to check clarity, grammar, and word usage. Any changes made based on this analysis was reviewed, considered, and adjusted by the authors and they take full responsibility for the content of the published article.